\documentclass{article}
\usepackage[utf8]{inputenc}
\usepackage{scicite}
\usepackage{ragged2e}
\usepackage{geometry}
\usepackage{url}
\usepackage{physabbrv}
\usepackage{hyperref}
\usepackage{graphicx}
\begin{document}
\noindent
\large{\textbf{Reply to the comment of T. Metcalfe and J. van Saders on the Science report "The Sun is less active than other solar-like stars" by T.~Reinhold, A.~I.~Shapiro, S.~K.~Solanki, B.~T. Montet, N.~A.~Krivova, R.~H.~Cameron, E.~M.~Amazo-G\'{o}mez}} \\[0.5cm]

Metcalfe \& van Saders \cite{comment} show that stars in the periodic sample of \cite{Reinhold2020} have on average smaller effective temperatures and slightly higher metallicities than stars in the non-periodic sample. This implies that the periodic stars have systematically smaller Rossby numbers than the non-periodic stars. \cite{comment} interpret this as a confirmation of their hypothesis of the evolutionary transition and decoupling between rotation and magnetic activity that stars experience above a critical Rossby number \cite{Metcalfe2016,Metcalfe2017}. According to their interpretation, the non-periodic stars and the Sun are either currently in transition to a magnetically inactive future or have already completed it, while the periodic stars did not yet start such a transition.

We agree with \cite{comment} that both the differences in the fundamental parameters, and the existence of the highly active periodic stars can be explained by the transition hypothesis, which was already indicated as a possible explanation of this phenomenon in \cite{Reinhold2020}. At the same time, we want to point out that the alternative explanation, that the Sun and other solar-like stars may occasionally experience epochs of high activity, also allows explaining the available data.

\begin{figure}[b]
  \centering
  \includegraphics[width=0.45\textwidth]{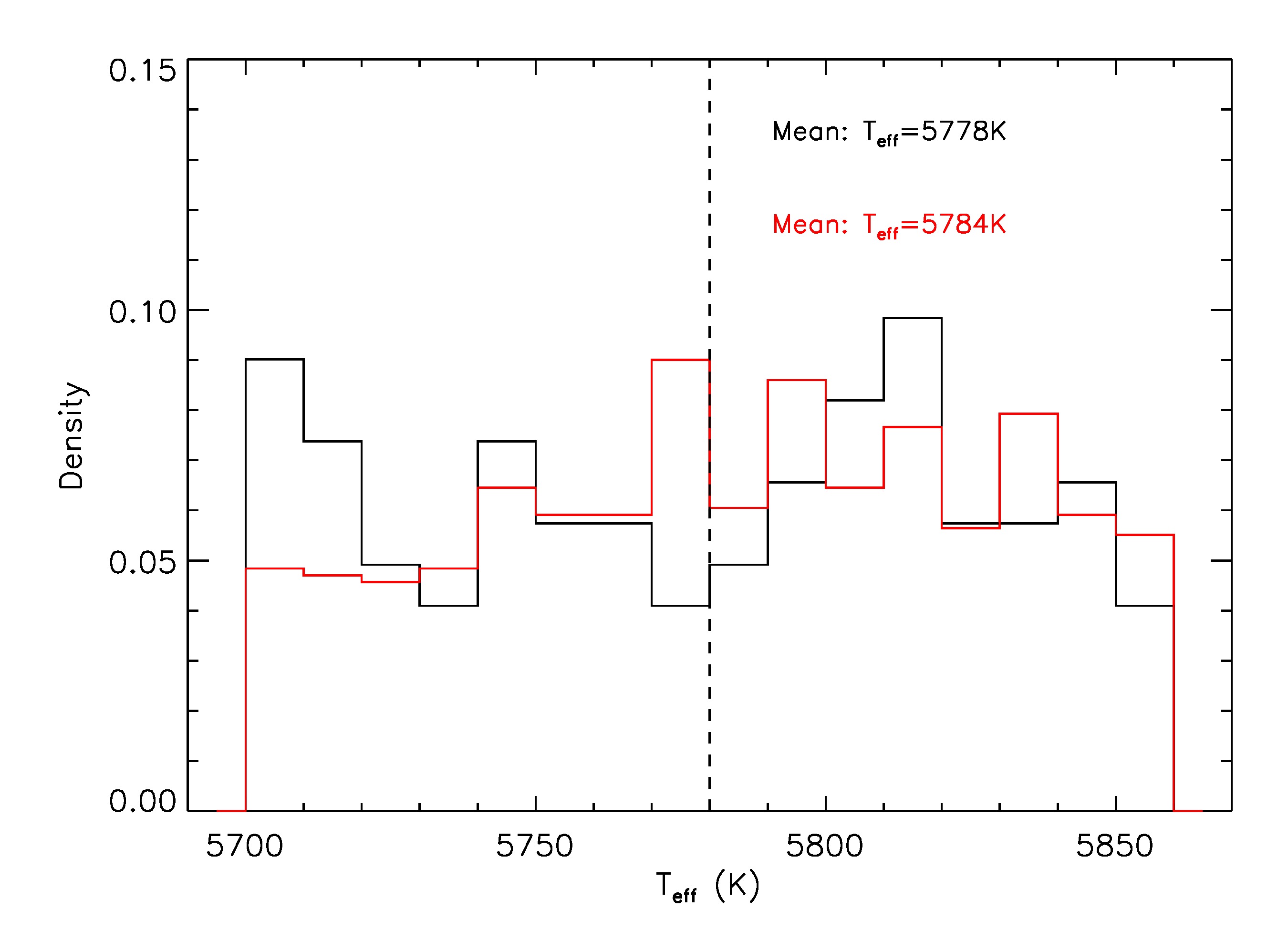}\includegraphics[width=0.45\textwidth]{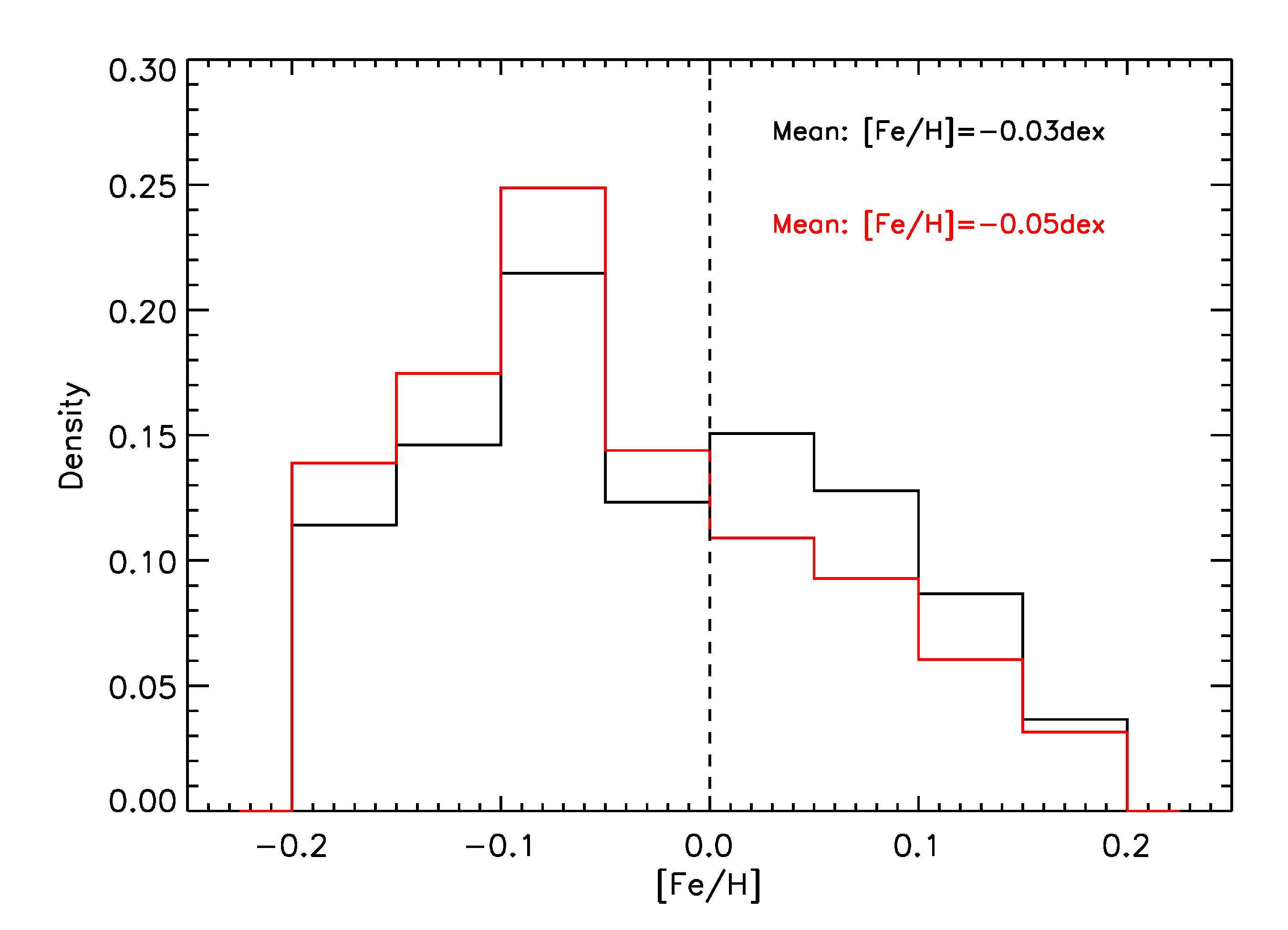}
  \caption{Left: Temperature distribution around the solar value (dashed line)
  of 122 periodic (black) and 744 non-periodic (red) stars, with mean values of 5778\,K, and 5784\,K, respectively. Right: Metallicity distribution around the solar value (dashed line) of 219 periodic (black) and 1174 non-periodic (red) stars, with mean values of -0.03~dex, and -0.05~dex, respectively.}
  \label{Teff_hist}
\end{figure}

In the original study, \cite{Reinhold2020} showed the variability distribution of stars with temperatures between 5500--6000\,K and metallicities from -0.8~dex to 0.3~dex. We note that  photometric effective temperatures and metallicities  have significant uncertainties (150--200\,K for effective temperature and about 0.2~dex for metallicity). Despite these uncertainties, the slightly different distribution of effective temperature and metallicities of the periodic and non-periodic stars in our original samples, pointed out by \cite{comment}, are real. Let us now consider a much narrower parameter range around the solar values (i.e. $T_{eff}=5780\pm80$\,K and $-0.2<[Fe/H]<0.2$~dex). In contrast to our original sample, both periodic and non-periodic stars show similar distributions of effective temperatures (left panel of Fig.~\ref{Teff_hist}) and of metallicities (right panel of Fig.~\ref{Teff_hist}). In other words, the trends reported by \cite{comment} are much less pronounced than in the original sample (either because they are indeed absent in such a domain of effective temperature and metallicities or because they are hidden by the uncertainties in photometric temperatures and metallicities). 

\begin{figure}[t]
  \centering
  \includegraphics[width=0.8\textwidth]{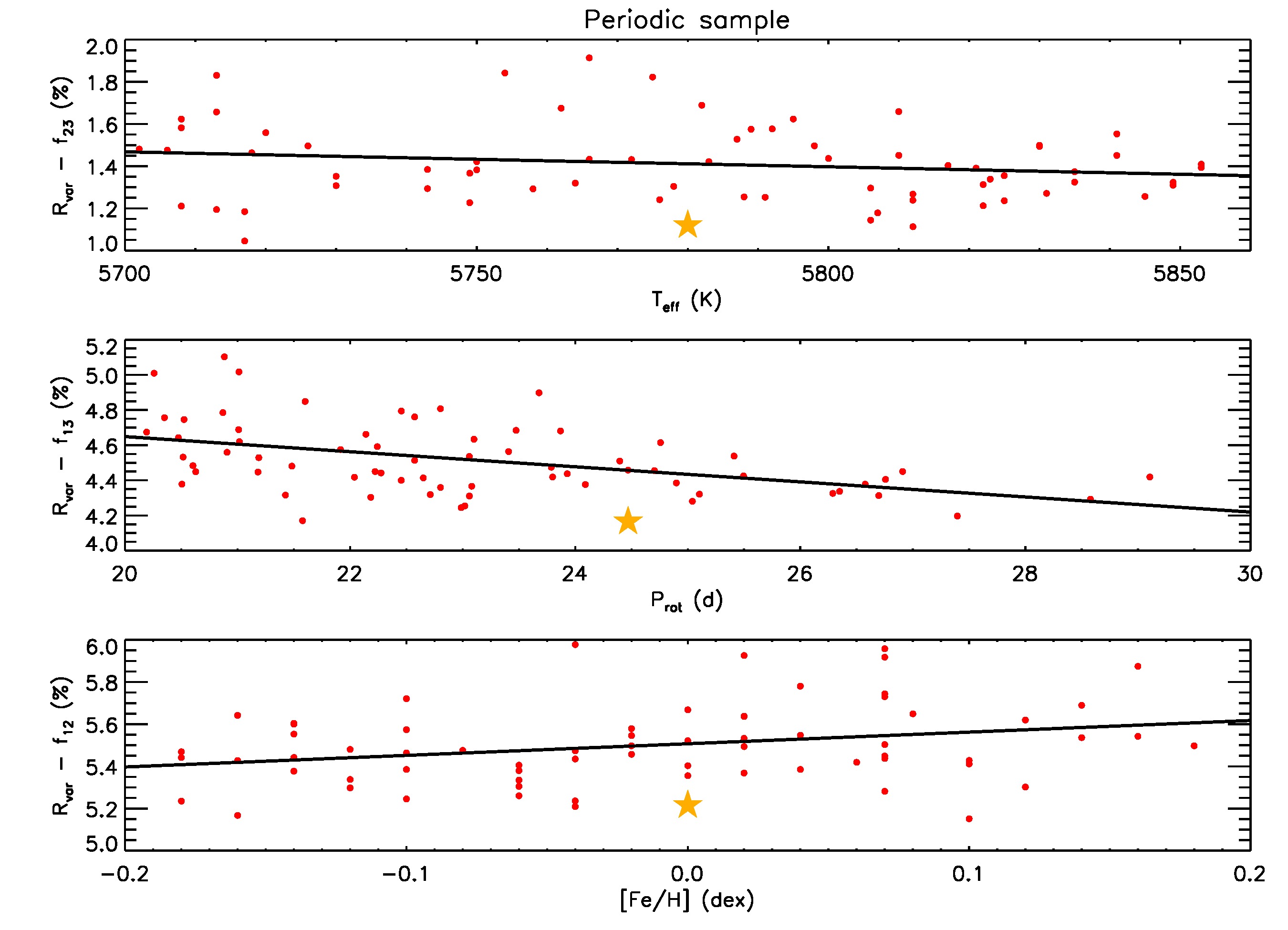}
  \caption{Dependence of $R_{var}$ on the fundamental parameters. The Sun is indicated by the yellow star symbol. The lines are from a multivariate regression. This figures shows the same as Fig.~S8 in \cite{Reinhold2020}, but for a narrower parameter range.}
  \label{correction}
\end{figure}

\begin{figure}[b]
  \centering
  \includegraphics[width=0.8\textwidth]{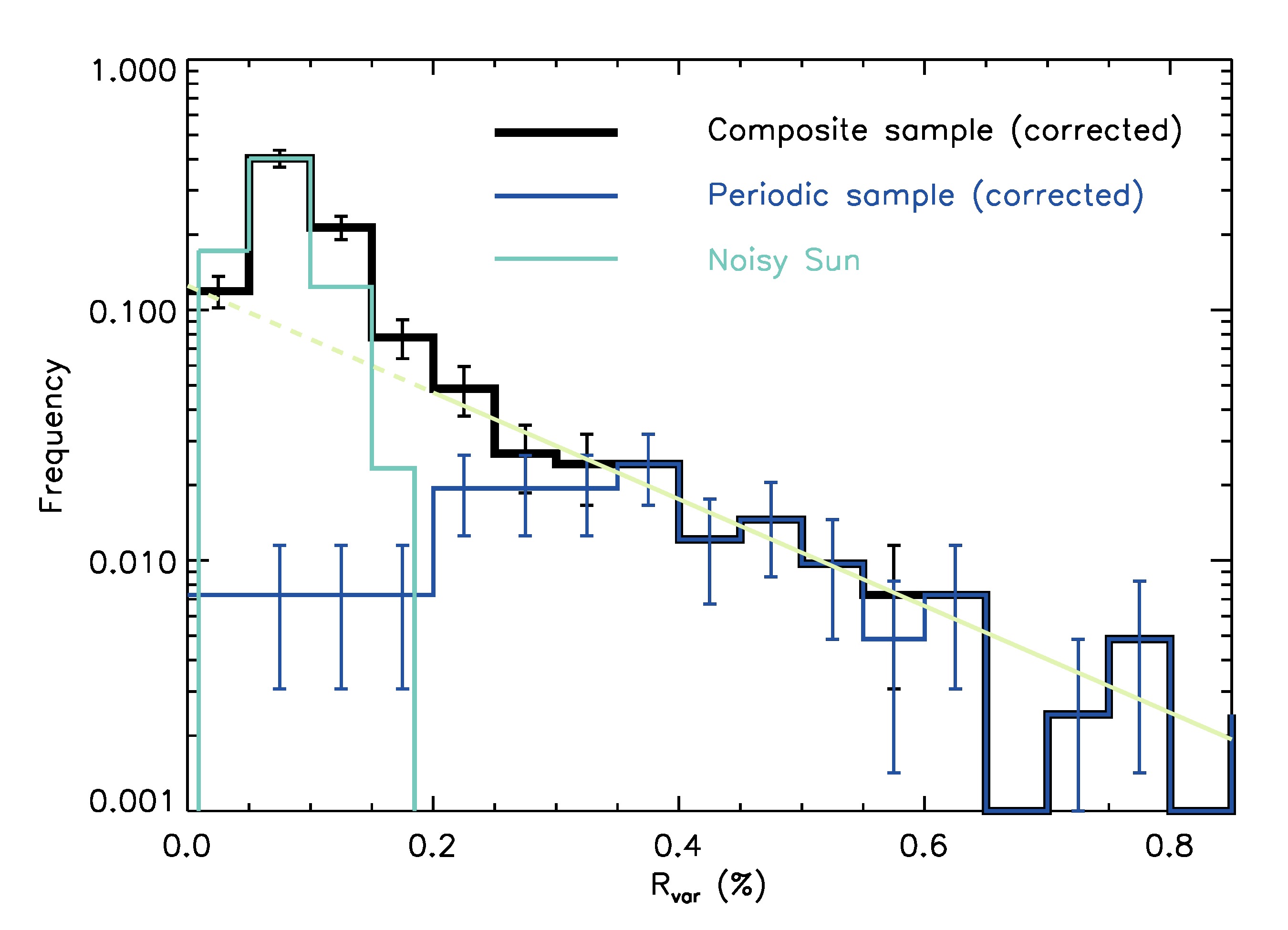}
  \caption{$R_{var}$ distribution of 67 periodic and 345 non-periodic stars with near-solar temperatures and metallicities.
  This figure shows the same as Fig.~3 in \cite{Reinhold2020}, but for a narrower parameter range.}
  \label{Rvar_dist}
\end{figure}

Interestingly, the ratio between periodic and non-periodic stars turns out to be almost the same as in the original distribution. Just the number of stars in the narrower sample is smaller: Considering only stars in these narrow parameter ranges reduces the original samples to 67 periodic and 345 non-periodic stars. Also the dependence of $R_{var}$ on the fundamental parameters remains almost unchanged (Fig.~\ref{correction}). Again, we correct for this dependence using a multivariate regression. Fig.~\ref{Rvar_dist} shows the variability distribution of stars in these narrow ranges. Qualitatively, we find a very similar distribution  as for the original sample (Fig.~3 in \cite{Reinhold2020}). In particular, the distribution shows that there are periodic stars populating the high variability tail that are otherwise nearly solar twins.

The dependence of the percentage of the periodic stars on effective temperature is amplified by stellar light curves becoming more regular with decreasing effective temperature \cite{Giles2017}. Such a change in the morphology of the light curves (which  \cite{Giles2017} attributed to the increase of the spot decay time towards cooler stars) leads to a better detectability of the rotation periods in cooler stars, and consequently, to the increase of the percentage of periodic stars with decreasing temperature. 

Another important factor one should keep in mind while analysing the trends in the percentage of the periodic stars is the dependence of the distribution of stellar rotation periods on effective temperature and metallicity. The efficiency of the spindown quite noticeably drops for stars hotter than the Sun, e.g. it becomes very weak for stars hotter than 6200\,K (the so-called Kraft break \cite{Kraft1967}). Also, the main-sequence lifetime decreases with effective temperature (remember that only main-sequence stars enter the sample of \cite{Reinhold2020}). Consequently, the number of stars with near-solar rotation periods (and, consequently, periodic stars in \cite{Reinhold2020}) significantly decreases for stars hotter than the Sun (see also Fig.~2 from \cite{McQuillan2014} which shows that the distribution of stellar rotation periods in periodic stars noticeably shifts to faster rotators with increasing effective temperature). 

Similarly, stars with lower metallicity but the same effective temperature have lower masses, and consequently, spend more time on the main sequence. Thus, for a fixed effective temperature, the number of old (e.g. older than the Sun) main-sequence stars is higher for stars with smaller metallicity. \cite{Amard2020} showed that while metallicity strongly affects the relationship between rotation period and age for stars {\it with fixed mass} the effect on stars with {\it fixed effective temperature} is much weaker. Consequently, older stars of a given effective temperature would generally rotate slower and most of them would be attributed to our non-periodic sample (since rotation periods get more difficult to detect for slow rotators). In other words, decreased metallicity leads to an increase of the pollution of our non-periodic sample by stars older and less active than the Sun. As a result, the fraction of periodic stars decreases with decreasing metallicity and the non-periodic stars on average have smaller metallicities, in agreement with \cite{comment}. 

In summary, we agree with the results of \cite{comment} that the ratio of periodic to non-periodic stars changes with effective temperature and metallicity, and the transition hypothesis allows explaining it. At the same time we argue that another explanation of this trend exists than the one favored by \cite{comment} and both hypotheses, which were already outlined in the original article \cite{Reinhold2020}, are consistent with the data.

\bibliography{biblothek}

\newcommand{\noop}[1]{}
\begin{thebibliography}{1}

\bibitem{comment}
T.~{Metcalfe}, J.~{van Saders}, {\it Comment on Science \textbf{368}, 518\/}
  (2020).

\bibitem{Reinhold2020}
T.~{Reinhold}, {\it et~al.\/}, {\it Science\/} {\bf 368}, 518 (2020).

\bibitem{Metcalfe2016}
T.~S. {Metcalfe}, R.~{Egeland}, J.~{van Saders}, {\it Astrophys. J. Lett.\/}
  {\bf 826}, L2 (2016).

\bibitem{Metcalfe2017}
T.~S. {Metcalfe}, J.~{van Saders}, {\it \solphys\/} {\bf 292}, 126 (2017).

\bibitem{Giles2017}
H.~A.~C. {Giles}, A.~{Collier Cameron}, R.~D. {Haywood}, {\it Mon. Not. R.
  Astron. Soc.\/} {\bf 472}, 1618 (2017).

\bibitem{Kraft1967}
R.~P. {Kraft}, {\it \apj\/} {\bf 150}, 551 (1967).

\bibitem{McQuillan2014}
A.~{McQuillan}, T.~{Mazeh}, S.~{Aigrain}, {\it Astrophys. J., Suppl. Ser.\/}
  {\bf 211}, 24 (2014).

\bibitem{Amard2020}
L.~{Amard}, S.~P. {Matt}, {\it \apj\/} {\bf 889}, 108 (2020).

\end{thebibliography}
\bibliographystyle{Science}

\end{document}